# Significant Reduction of the Microwave Surface Resistance of $MgB_2$ Films by Surface Ion Milling

Sang Young Lee,[a)*] J. H. Lee,[a)] Jung Hun Lee,[a)] J. S. Ryu,[a)] J. Lim,[a)] S. H. Moon,[b)] H. N. Lee,[b)], H. G. Kim,[b)] and B. Oh[b)]

[a)] Department of Physics, Konkuk University, Seoul 143-701, Korea
[b)] LG Electronics Institute of Technology, Seoul 137-724, Korea

The microwave surface resistance $R_S$ of $MgB_2$ films with the zero-resistance temperature of ~ 39 K was measured at 8.0 – 8.5 GHz. The $MgB_2$ films were prepared by deposition of boron films on $c$-cut sapphire, followed by annealing in a magnesium vapor environment. The $R_S$ appeared significantly reduced by ion milling of the as-grown $MgB_2$ film surface, with the observed $R_S$ of ~ 0.8 mΩ at 24 K for an ion-milled $MgB_2$ film as small as 1/15 of the value for the corresponding as-grown $MgB_2$ film. The reduced $R_S$ of the ion-milled $MgB_2$ films is attributed to the effects of the Mg-rich metallic layer existing at the surfaces of the as-grown $MgB_2$ films.

High critical current density ($J_C$) observed in $MgB_2$ with the critical temperature ($T_C$) of ~ 39 K provides strong possibilities that the microwave surface resistance ($R_S$) of $MgB_2$ could be low enough at temperatures obtainable using close-cycle cryostats[1-6]. Reports on the microwave properties of polycrystalline $MgB_2$ pellets and wires have recently been made, where the surface resistance of polycrystalline $MgB_2$ wires appeared to be comparable with that of single crystal $YBa_2Cu_3O_{7-\delta}$ (YBCO) at 20 K with the value still much above the estimated value for a pure Bardeen-Cooper-Schrieffer superconductor[7, 8]. With the microwave properties of $MgB_2$ single crystals unknown despite reports on their fabrications[9], use of 39 K $MgB_2$ films could provide an alternative way for understanding the intrinsic microwave properties of $MgB_2$. However, microwave properties of 39 K $MgB_2$ films, which may have great potential for microwave applications, are yet to be investigated.

$MgB_2$ films have been fabricated on different substrates such as $r$-cut and $c$-cut sapphire and MgO using laser ablation or electron beam evaporation [5, 6]. To date, most $MgB_2$ films have been prepared *ex situ* with the stoichiometric composition realized by reaction of Mg vapor with a boron layer inside a quartz tube. As a result, the surface of *ex-situ* prepared $MgB_2$ films appears relatively rough, making it difficult to prepare $MgB_2$ Josephson junctions. Furthermore, it is very likely that the composition of the surface layer could be different from the stoichiometric $MgB_2$, with Mg-rich phases including Mg and MgO existing at the surface.

In this letter, we report on the microwave $R_S$ of two $MgB_2$ films with the zero-resistance temperature ($T_{C,zero}$) of ~ 39 K prepared under different deposition conditions. Especially, the effects of the metallic surface layer on the microwave $R_S$ of $MgB_2$ films were studied here. Significant reduction of the microwave $R_S$ by 15 times could be observed after ion milling of the film surface layer. Attempts are made to explain the reduction of the $R_S$ based on the effects of a metallic layer at the surface of the $MgB_2$ films.

Two $MgB_2$ films were prepared on $c$-cut sapphire by depositing boron films on the substrates, followed by annealing in a Mg vapor environment inside quartz tubes. One $MgB_2$ film ($MgB_2$-1A) was prepared by annealing at 825 °C for 20 minutes, while the other ($MgB_2$-2A) was annealed at 800 °C for 30 minutes. The dimensions of the $MgB_2$ films are 11 x 11 mm$^2$ and 5 x 9 mm$^2$, respectively, with the thickness of ~ 420 nm. More details of the sample preparation have been reported elsewhere[6].

The films appeared to be single phase from the X-ray diffraction (XRD) data. The dc resistance data showed that the films had the critical onset temperature of ~ 39.2 K and the transition width of ~ 0.3 K, as displayed in Fig. 1. Auger spectroscopy study revealed that the surface layers of our typical $MgB_2$ films are Mg-rich up to the depth of ~ 50 nm. The $R_S$ of $MgB_2$-1A and $MgB_2$-2A were measured using an open-ended $TE_{01\delta}$ mode cavity resonator made of oxygen-free high conductivity copper (OFHC) loaded with a rutile-phase $TiO_2$ (henceforth called 'rutile') rod. For measurements of the $R_S$ of the $MgB_2$ film, a YBCO film, which was used as a reference, was placed at the bottom of the cavity with the $MgB_2$ film placed at the top. Later the as-grown MgB$_2$-1A and MgB$_2$-2A films were ion-milled for 10 minutes, after which they were designated MgB$_2$-1I and MgB$_2$-2I, and then remeasured. It is noted that 10 minutes' ion-milling reduced the film thickness by ~ 55 nm.

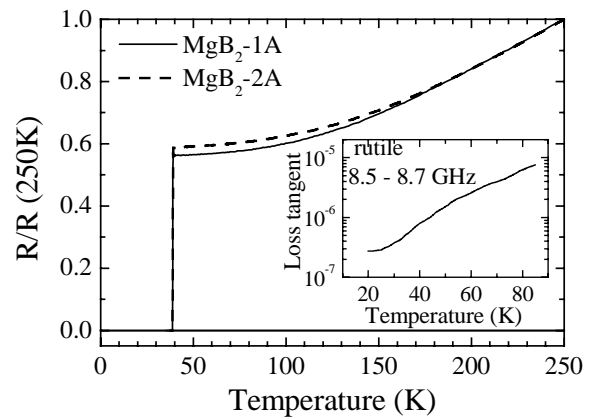

Fig. 1. The resistance ($R$) data as a function of temperature normalized to the resistance at 250 K ($R$(250K)) for the as-grown $MgB_2$-1A and $MgB_2$-2A. Both films appear to have $T_{C,zero}$ ~ 39 K with the transition width of ~ 0.3 K. Inset: The temperature dependence of the loss tangent of the rutile used for this experiment.

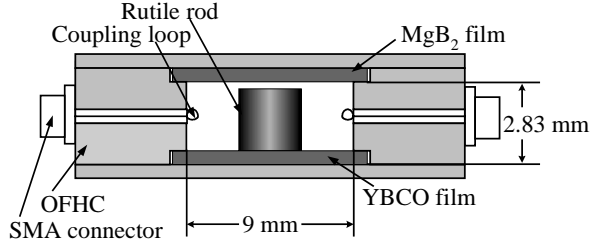

Fig. 2. A diagram of the TE$_{01\delta}$ mode rutile-loaded cavity resonator. A YBCO film placed at the bottom was used as a reference. In measurements of the $R_S$ of MgB$_2$-2A and MgB$_2$-2I, a 0.2 mm-thick OFHC disk with a 4.5 mm-in-diameter hole at the center was used as the top endplate of the cavity with MgB$_2$-2A and MgB$_2$-2I placed right above the hole.

Figure 2 shows a diagram of the TE$_{01\delta}$ mode rutile-loaded cavity resonator, which was used for measurements of the $R_S$ of the MgB$_2$ films. Our measurement system, based on a HP8510C network analyzer and a close-cycle cooler, is computer-controlled with a temperature stability of ± 0.2 K. The dimensions of the rutile rod are 3.88 x 2.73 mm$^2$ for the diameter and height, respectively. The gap between the rutile rod and the top plate was set to 0.1 mm. In measurements of the $R_S$ of MgB$_2$-2A and MgB$_2$-2I, a polished 0.2 mm-thick OFHC disk with a 4.5 mm-in-diameter central hole was used as the top endplate of the cavity with MgB$_2$-2A and MgB$_2$-2I placed right above the hole. The TE$_{01\delta}$ mode resonant frequency ($f_0$) of the resonator appeared to be ~ 8.5 GHz with MgB$_2$-1A (and MgB$_2$-1I) and ~ 8.0 GHz with MgB$_2$-2A (and MgB$_2$-2I) at temperatures of 20 - 40 K. During measurement, care was taken to couple the resonator weakly to microwave power with the insertion loss higher than 30 dB. In this case, the differences between the loaded $Q$ ($Q_L$) and the unloaded $Q$ ($Q_0$) are within 5 %. The $R_S$ of the MgB$_2$ films was calculated using the following equation for the rutile-loaded resonator with a YBCO film at the bottom and a MgB$_2$ film at the top, $1/Q_0 = R_S^{MgB2}/\Gamma_T + R_S^{MgB2}/\Gamma_B + R_S^{Cu}/\Gamma_S + k \cdot \tan\delta$ for MgB$_2$-1A and MgB$_2$-1I, with $R_S^{MgB2}$ and $R_S^{YBCO}$ denote the $R_S$ of the MgB$_2$ and the YBCO films, $R_S^{Cu}$, the $R_S$ of the OFHC side wall, $\Gamma_T$, $\Gamma_B$, and $\Gamma_S$, the respective geometric factors of the top and the bottom endplates, and the cavity side wall, $k$, the filling factor, and $\tan\delta$, the loss tangent of rutile. The calculated geometric factors and the filling factor are $\Gamma_B = 217$ Ω, $\Gamma_T = 219$ Ω, $\Gamma_S = 26024$ Ω and $k = 0.99773$. In measurements of the $R_S$ of MgB$_2$-2A and MgB$_2$-2I, $R_S^{MgB2}/\Gamma_T$ in the expression for $1/Q_0$ was replaced by $R_S^{MgB2}/\Gamma_{T1} + R_S^{Cu}/\Gamma_{T2}$ with $\Gamma_{T1} = 242$ Ω and $\Gamma_{T2} = 2301$ Ω.

The $R_S$ of the YBCO film placed at the bottom of the cavity was measured using the measured $Q_0$ of a TE$_{01\delta}$ mode sapphire-loaded cavity resonator with $f_0 = 19.5$ GHz [10]. Also, $\tan\delta$ of rutile was separately measured by comparing the $Q_0$ of a sapphire-loaded cavity resonator (TE$_{01\delta}$ mode $f_0$ ~ 19.5 GHz) and a rutile-loaded cavity resonator (TE$_{01\delta}$ mode $f_0$ ~ 8.5 GHz) having the same YBCO endplates. We assumed $R_S \propto f^2$ to obtain the $R_S$ of the YBCO films at ~ 8.5 GHz and ~ 8.0 GHz. For the loss tangent of sapphire as a function of temperature, the data reported by Krupka et al.

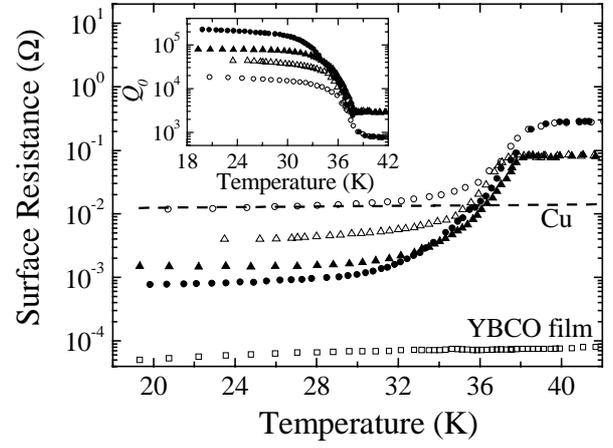

Fig. 3. The temperature dependence of the $R_S$ of the as-grown MgB$_2$ films and the ion-milled MgB$_2$ films. Note that the data for MgB$_2$-1A (open circle) and MgB$_2$-1I (filled circle) were taken at ~ 8.5 GHz while the data for MgB$_2$-2A (open triangle) and MgB$_2$-2I (filled triangle) were taken at ~ 8.0 GHz. The values of $R_S$ for OFHC (dotted line) and a YBCO film (square) are shown for comparison. Inset: The TE$_{01\delta}$ mode $Q_0$ of the rutile-loaded resonator as a function of temperature with each of the MgB$_2$ films placed at the top.

were used [11]. Two identical YBCO films used for measurements of $\tan\delta$ of rutile were grown on CeO$_2$-buffered sapphire by co-evaporation. The sapphire and the rutile rods were provided by Crystec. In the inset of Fig. 1, we show the loss tangent of rutile as a function of temperature at 8.5 – 8.7 GHz.

Figure 3 shows the $R_S$ as a function of temperature for MgB$_2$-1A MgB$_2$-1I, MgB$_2$-2A MgB$_2$-2I and OFHC, respectively. In the figure, we see that the $R_S$ of MgB$_2$-1A appeared higher than that of OFHC even at temperatures below $T_C$. However, the $R_S$ of the ion-milled MgB$_2$-1A (i.e., MgB$_2$-1I) appeared significantly reduced with $R_S \approx 0.8$ mΩ at 8.5GHz at 24 K, which is much lower than that of MgB$_2$-1A and OFHC. Also, the $R_S$ of the ion-milled MgB$_2$-2A (i.e., MgB$_2$-2I) appeared significantly reduced compared to the corresponding value of MgB$_2$-2A, with $R_S \approx 1.5$ mΩ at 8.0 GHz at 24 K. These values are only ~ 1/15 and ~ 1/3 of the corresponding values for MgB$_2$-1A and MgB$_2$-2A, respectively. We note in Fig. 3 that the reduction of the $R_S$ due to the surface ion milling is 11 mΩ and 2.5 mΩ at 24 K for MgB$_2$-1A and MgB$_2$-2A, respectively, with the corresponding values slowly increasing to 16 mΩ and 4 mΩ up to 34 K. The temperature dependences of the $Q_0$ used for calculating the $R_S$ are shown in the inset of Fig. 3.

In Fig. 4, the $R_S$ data of MgB$_2$-1I and MgB$_2$-2I at 10 GHz, calculated under the assumption of $R_S \propto f^2$, are compared with a YBCO film as a function of the reduced temperature $T/T_C$. The reported values for the $R_S$ of MgB$_2$ wire [7] are also shown for comparison. Interestingly, the $R_S$ of MgB$_2$-1I appeared almost the same with that of the MgB$_2$ wire at temperatures below 35 K as seen in Fig. 4, although the normal state $R_S \approx 8.7$ mΩ at $T_C$ for MgB$_2$-1I appeared about 10 times higher than that of the MgB$_2$ wire. It is also noted that the observed $R_S$ of our ion-milled MgB$_2$ films appeared about 10 times higher than that of epitaxially grown YBCO films for $T/T_C < 0.6$

Our results suggest that the observed high $R_S$ of the as-grown MgB$_2$ films can be attributed to the existence of non-stoichiometric composition at the surface of the MgB$_2$ films.

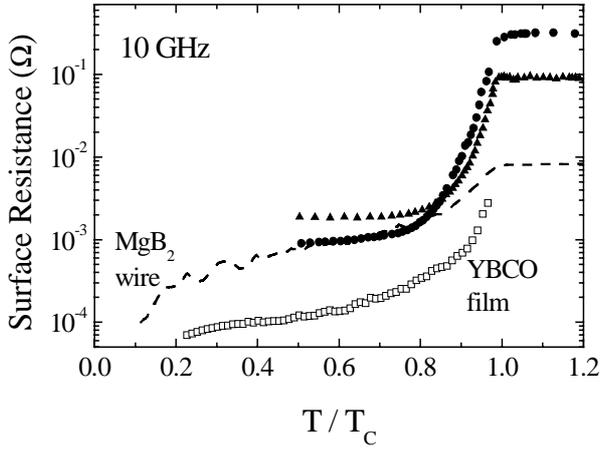

Fig. 4. The $R_S$ of MgB$_2$-1I (circle), MgB$_2$-2I (triangle) and a YBCO film (square) as a function of the reduced temperature $T/T_C$ at 10 GHz. $T_C$ of 39 K and 88 K were used for the MgB$_2$ films and the YBCO film, respectively. The $R_S$ at 10 GHz was obtained from the data in Fig. 3 under the assumption of $R_S \propto f^2$ at $T < T_C$ and $R_S \propto f^{1/2}$ for $T > T_C$. The reported $R_S$ of MgB$_2$ wire[7] (dotted line) is also shown for comparison.

Since the Mg-rich surfaces of MgB$_2$-1A and MgB$_2$-2A are believed to be metallic from the resistance data (see Fig. 1), there can be significant effects of the Mg-rich metallic layer on the effective $R_S$ of the MgB$_2$-1A and MgB$_2$-2A. The effective surface impedance $Z_S^{eff}$ of a superconductor with a single metallic surface layer of thickness $d$ and the skin depth $\delta$ can be described by $Z_S^{eff} = Z_C \cdot [Z_C \tanh(\gamma d) + Z_S]/[Z_C + Z_S \tanh(\gamma d)]$ [12]. Here $Z_C$ (= $R_C (1 + i)$) and $Z_S$ (= $R_S + iX_S$) denote the surface impedance of the metallic layer and the superconductor, respectively, with $R_C$ denoting the surface resistance of the metallic layer, $\gamma$ (= $(1 + i)/\delta$), the propagation constant inside the metallic layer, and $R_S$ and $X_S$, the surface resistance and the surface reactance of the superconductor, respectively. Now, if we assume that a single 55 nm-thick Mg layer is placed on top of MgB$_2$-1I and MgB$_2$-2I and that $\lambda_0 \approx 140$ nm with the temperature dependence of $1/\lambda^2 = (1/\lambda_0^2) \cdot [1 - (T/T_C)^4]^{1/2}$ assumed for MgB$_2$[13], the calculated values $\Delta R_S^{CAL}$ for the enhanced $R_S$ due to the Mg layer (i.e., the difference between the real part of $Z_S^{eff}$ and the $R_S$) would be ~ 3.4 mΩ and ~ 2.6 mΩ for MgB$_2$-1I and MgB$_2$-2I, respectively at 24 K. For the calculation, we used $R_C = \omega\mu_0\delta/2$ and $X_S = \omega\mu_0\lambda$ with $\omega = 2\pi f$, $\mu_0$ denoting the permeability of the vacuum, $T_C = 39$ K, $\delta = (\omega\mu_0/2\rho) = 85$ nm[14]. It is noted that the enhanced $R_S$ of 2.6 mΩ calculated at 8.0 GHz at 24 K is almost the same as the experimental difference $\Delta R_S^{EXP}$ of 2.5 mΩ in the $R_S$ of MgB$_2$-2A and MgB$_2$-2I at temperatures below 26 K. Furthermore, $\Delta R_S^{CAL}$ appeared close to $\Delta R_S^{EXP}$ in the $R_S$ of MgB$_2$-2A and MgB$_2$-2I up to 34 K, with $\Delta R_S^{CAL}$ of 3.5 mΩ well compared with $\Delta R_S^{EXP}$ of 4.0 mΩ at 34 K. However, the respective $\Delta R_S^{EXP}$ of ~ 11 mΩ and ~ 15 mΩ at 8.5 GHz at 24 K and 34 K in the $R_S$ of MgB$_2$-1A and MgB$_2$-1I appeared much larger than the corresponding $\Delta R_S^{CAL}$ of 3.4 mΩ and 3.6 mΩ, suggesting that the surface layer of MgB$_2$-1A cannot be regarded as a single Mg layer. $\Delta R_S^{CAL}$ could be as somewhat larger for MgB$_2$-1I if its $\lambda_0$ was larger than 140 nm, which, however, seems not the case here. Considering the different annealing temperature and time used in preparing MgB$_2$-1A and MgB$_2$-2A, the characteristics of the surface layer of MgB$_2$-1A are plausibly different from those of MgB$_2$-2A. Further studies are needed to clarify the correct reasons for the observed significant reduction of the $R_S$ after surface ion-milling.

Summarizing, two MgB$_2$ films, MgB$_2$-1A and MgB$_2$-2A, with $T_{C,zero}$ ~ 39 K were prepared on c-cut sapphire by reaction of Mg vapor with boron films at 825 °C for 20 minutes and at 800 °C for 30 minutes, respectively. The $R_S$ of both MgB$_2$-1A and MgB$_2$-2A appeared significantly reduced after ion-milling of the surface by ~ 55 nm, with the $R_S$ reduced by as much as 15 times at 24 K. The observed high $R_S$ of MgB$_2$-1A and MgB$_2$-2A is attributed to the existence of a Mg-rich metallic surface layer on the as-grown MgB$_2$ films, with the respective changes of 2.5 mΩ and 4 mΩ in the $R_S$ of MgB$_2$-2A at 24 K comparable to the corresponding enhanced $R_S$ of 2.6 mΩ and 3.5 mΩ calculated under the assumption of the existence of a single 55 nm-thick Mg layer. However, respective values of only ~ 1/3 and ~ 1/4 of the change in $R_S$ could be explained for MgB$_2$-1A at 24 K and 34 K within this context, implying that the Mg-rich surface layer can affect the $R_S$ of MgB$_2$ films in a complicated way depending on their growth conditions.

The authors express sincere thanks to Dr. M. Hein for valuable comments on the effect of metallic surface layer and acknowledge technical assistance of Mr. J. D. Park and K. Y. Lee.

This work has been supported by Korean Ministry of Science and Technology and the National Research Laboratory project.